\documentclass[twocolumn,preprintnumbers,amsmath,aps] {revtex4}
\usepackage{graphicx}
\usepackage{dcolumn}
\usepackage{bm}
\usepackage{color}
\usepackage{microtype}
\usepackage{multirow}
\usepackage{booktabs}
\begin{document}

\title{The critical field and specific heat in the electron- and hole-doped graphene superconductors}


\author{Ewa A. Drzazga-Szcz{\c{e}}{\'s}niak$^{1}$}\email{ewa.drzazga@pcz.pl}
\author{Adam Z. Kaczmarek$^{2}$}


\affiliation{$^1$Department of Physics, Faculty of Production Engineering and Materials Technology, Cz{\c{e}}stochowa University of Technology, 19 Armii Krajowej Ave., 42200 Cz{\c{e}}stochowa, Poland}
\affiliation{$^2$Department of Theoretical Physics, Faculty of Science and Technology, Jan D{\l}ugosz University in Cz{\c{e}}stochowa, 13/15 Armii Krajowej Ave., 42200 Cz{\c{e}}stochowa, Poland}


\date{\today}


\begin{abstract}
Doping is one of the most prominent techniques to alter properties of a given material. Herein, the influence of the electron- and hole-doping on the selected superconducting properties of graphene are considered. In details, the Migdal-Eliashberg formalism is employed to analyze the specific heat and the critical magnetic field in the representative case of graphene doped with nitrogen or boron, respectively. It is found that the electron doping is much more favorable in terms of enhancing the aforementioned properties than its hole counterpart. These findings are appropriately summarized by the means of the dimensionless thermodynamic ratios, familiar in the Bardeen-Cooper-Schrieffer theory. To this end, the perspectives for future research on superconductivity in graphene are drawn.
\end{abstract}

\maketitle
\vspace{0.5cm}

\section{Introduction}

The two-dimensional carbon allotrope known as graphene became one of the most important materials in nanoscience due to its wide range of intriguing properties \cite{novoselov2004,das2008,Ye2010}. In particular, a lot of effort was devoted to the exploration of superconductivity in a various graphene-based structures \cite{Profeta2012,salvini2010,zhou2015,blackschaffer2007,kiesel2012,ma2014,ludbrook2015}. Since superconductivity in pure graphene is absent, doping it with non-carbon elements turned out to be an efficient approach toward induction of the discussed phase \cite{Profeta2012,salvini2010,zhou2015, szczesniak2019, szczesniak2021}. Such promising behavior led to many considerations over the recent years. In general, graphene modifications were proposed to induce not only conventional but also unconventional superconductivity \cite{blackschaffer2007,kiesel2012,ma2014,ludbrook2015}, with the former phase being more anticipated because of its strong theoretical foundations. Strictly speaking, the conventional (phonon-mediated) superconductivity can be induced in graphene by increasing the electron-phonon coupling parameter ($\lambda$). This can be done by doping graphene with electrons or holes. A prominent example of such process are the lithium-decorated graphene \cite{Profeta2012}, graphane \cite{salvini2010} or the nitrogen/boron-doped graphene \cite{zhou2015}.

In the context of the above developments, the nitrogen/boron-doped graphene \cite{zhou2015} appears as an exemplary material to study low-dimensional superconducting phase of interest. This is due to the fact that both of the mentioned materials exhibit relatively high superconducting properties and allows comparing the two doping strategies on the same footing. Therefore, we attempt to investigate the hitherto not discussed thermodynamic properties of the boron and nitrogen doped graphene. In particular, we use the Eliashberg formalism to analyze the critical magnetic field and the specific heat. Our work is organized as follows: in Sec. II we describe theoretical model of the Migdal-Eliashberg equations. Next, in Sec. III we discuss the specific heat and related quantities obtained from the numerical analysis of the theoretical model. This study is concluded by summary and remarks regarding future perspectives in Sec. IV.

\section{Theoretical model}

\begin{center}
\begin{figure*}[ht!]
\includegraphics[width=\textwidth]{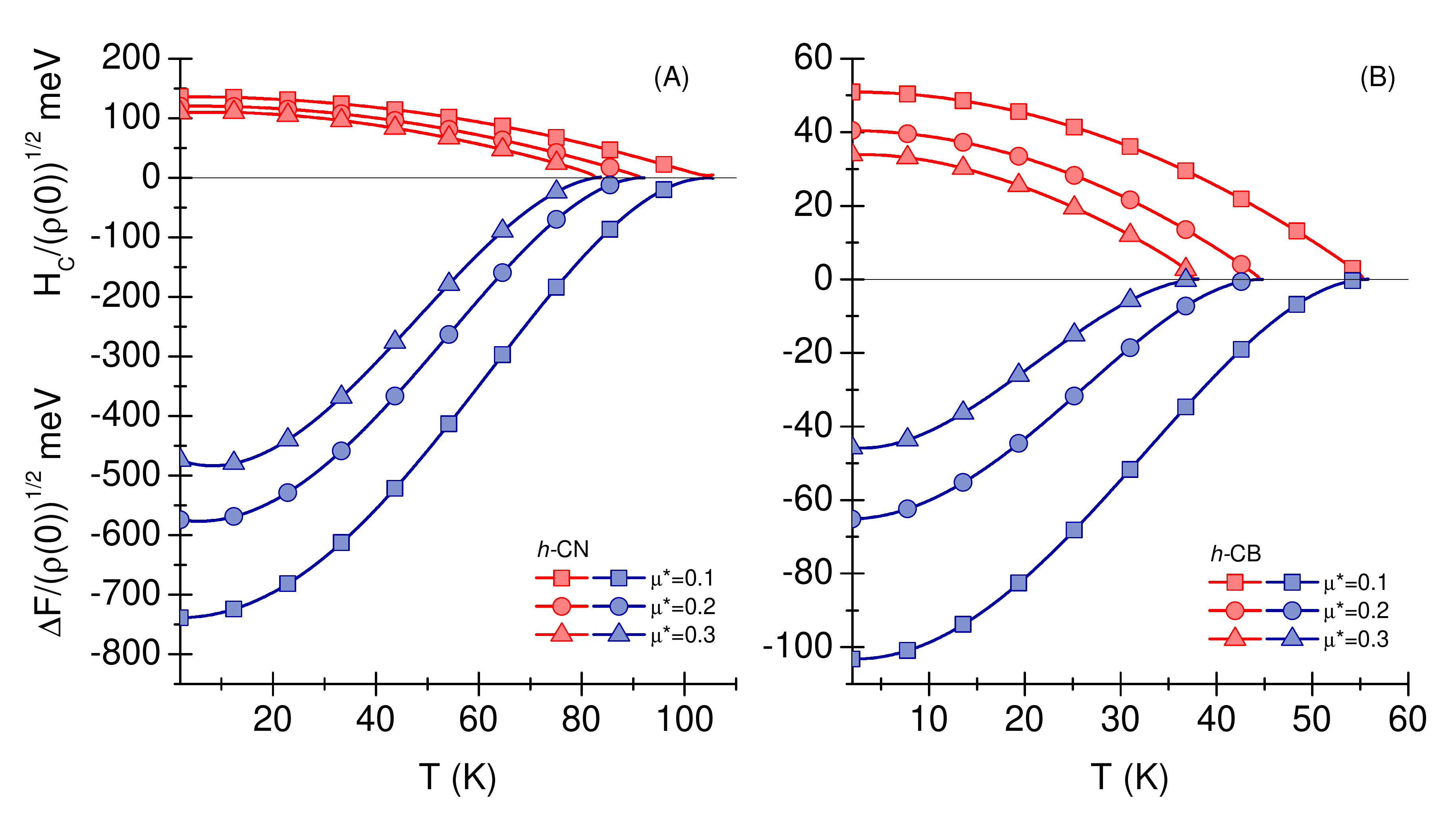}
\caption{The free energy difference (lower panels) and hte critical magnetic field (upper panels) as a function of temperature for the selected $\mu^{*}$ values in (A) the electron- ($h-$CN) and (B) hole-doped graphene ($h-$CB).}
\label{f1}
\end{figure*}
\end{center}

In order to describe the thermodynamic properties of the electron- ($h-$CN) and hole-doped ($h-$CB) graphene, we adopt the isotropic approximation of the Migdal-Eliashberg equations \cite{migdal1958, eliashberg1960, carbotte1990}. From the obtained solutions of the Migdal-Eliashberg equations on the imaginary axis, we are able to determine the order parameter ($\Delta_{n}=\Delta\left(i\omega_{n}\right)$) with the associated wave function renormalization factor ($Z_{n}=Z\left(i\omega_{n}\right)$), given as:
\begin{eqnarray}
\Delta_{n}Z_{n} &=& \frac{\pi}{\beta}\\ \nonumber
&\times& \sum_{m=-M}^{M} \frac{K\left(i\omega_{n}-i\omega_{m}\right)-\mu^{\star}\theta\left(\omega_{c}-|\omega_{m}|\right)}
{\sqrt{\omega_m^2Z^{2}_{m}+\Delta_{m}^{2}}}\Delta_{m}^2,
    \label{eq1}
\end{eqnarray}
and
\begin{equation}
Z_{n}=1+\frac{1}{\omega_{n}}\frac{\pi}{\beta}\sum_{m=-M}^{M}
\frac{K\left(i\omega_{n}-i\omega_{m}\right)}{\sqrt{\omega_m^2Z^{2}_{m}+\Delta^{2}_{m}}}\omega_{m}Z_{m},
    \label{eq2}
\end{equation}
where $\beta=1\slash k_{B}T$ denotes inverse temperature, with $k_{B}$ being the Boltzmann constant ($k_{B}$). In what follows, the Matsubara frequency can be written as $\omega_{n}=\left(\pi\slash\beta\right)\left(2n-1\right)$. Moreover, $K\left(z\right)=2\int_0^{\omega_{\rm{max}}} d \left( \alpha^{2}F(\omega)\omega \right) \left[(\omega_{n}-\omega_{m})^2+\omega^2 \right]$ and stands for the electron-phonon paring kernel. Note that the $\alpha^{2}F\left(\omega\right)$ function, conventionally refereed to as the Eliashberg functions, is adopted from the study of Zhou {\it et al.} \cite{zhou2015}. Therein, this function was calculated for all cases of interest within the density function theory, as implemented in the Quantum-ESPRESSO package. To this end, in Eq. (\ref{eq1}), the electron-electron depairing interactions are modeled by the Coulomb pseudopotential ($\mu^{\star}$) parameter, defined as $\mu^{\star}\equiv\mu^{\star}\theta(\omega_{c}-|\omega_{m}|)$, where $\theta$ is the Heaviside function.

The introduced Eliashberg equations are solved here by using the self-consistent iterative procedures developed previously in \cite{szczesniak2006}. The stability of the numerical procedures is reached at around the 2201 Matsubara frequencies, assuming $T_{0}=2$ K and the phonon frequency cut-off ($\omega_{c}$) equal to $10\omega_{max}$, where  $\omega_{\rm{max}}=132.6$ meV ($h-$CN) and $\omega_{\rm{max}}=124.5$ meV ($h-$CB) and denotes the maximum value of the phonon frequency defined by the adopted $\alpha^{2}F\left(\omega\right)$ function. Finally, $\mu^{\star}$ is set to 0.1-0.3, as suggested by Ashcroft \cite{ashcroft2004}.

\begin{center}
\begin{figure*}[ht!]
\includegraphics[width=\textwidth]{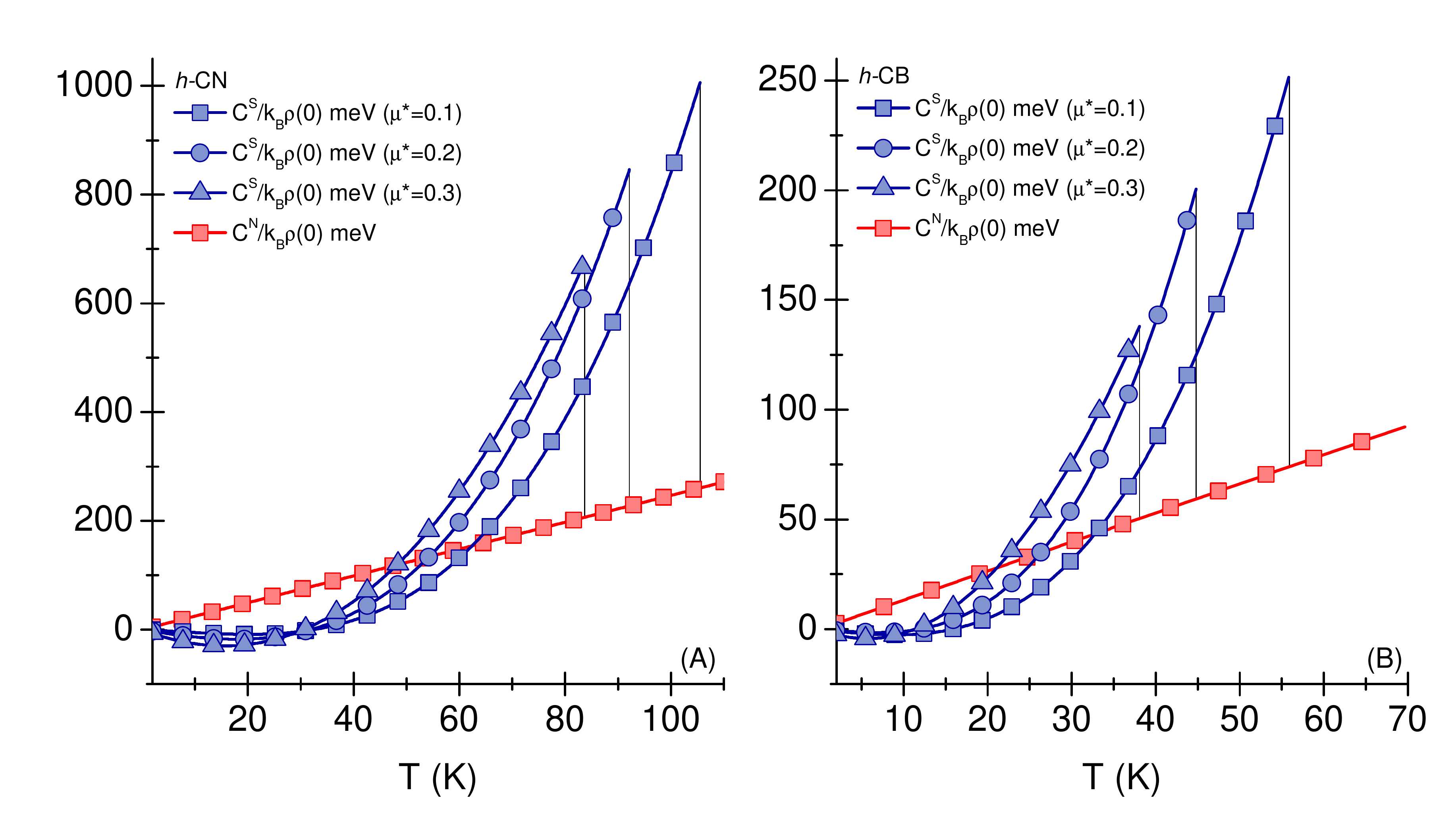}
\caption{The specific heat for the superconducting and normal state as a function of the temperature for the selected $\mu^{*}$ values in (A) the electron- ($h-$CN) and (B) hole-doped graphene ($h-$CB). The specific heat jump at the critical temperature is marked by the solid vertical line.}
\label{f2}
\end{figure*}
\end{center}

\section{Results and discussion}

We are starting the description of the selected thermodynamic characteristics by recalling the free energy difference between the normal and superconducting state ($\Delta F$):
\begin{align}\nonumber
    \frac{\Delta F}{\rho(0)}=-\frac{2\pi}{\beta} \sum_{m=1}^M(\sqrt{\omega^2_m+ \Delta_m^2}-|\omega_m|) \\ \times \Big(Z_m^{S}-Z_m^{N} \frac{|\omega_m|}{\sqrt{\omega_m^2+\Delta^2_m}}\Big),
    \label{eq3}
\end{align}
with the renormalization factors $Z_m^S$ and $Z_m^N$ for the superconducting ($S$) and normal ($N$) state. In the bottom panel of the Figs. (\ref{f1}) we have presented the dependence of the $\frac{\Delta F}{\rho(0)}$ on the temperature for both considered structures. We remark that the negative values of the considered function confirm the thermodynamic stability of the superconducting phase for the $h$-CN (A) and $h$-CB (B). Moreover, it is observed that the increase of the Coulomb pseudopotential leads to the decrease of the $\Delta F(0)$ parameter, since:
\begin{align}
[\Delta F(0)]_{\mu^*=0.3}\slash [\Delta F(0)]_{\mu^*=0.1}\approx 0.63,   
    \label{eq4}
\end{align}
and
\begin{align}
    [\Delta F(0)]_{\mu^*=0.3}\slash [\Delta F(0)]_{\mu^*=0.1}\approx 0.44.
    \label{eq5}
\end{align}
for the $h-$CN and $h-$CB, respectively (assuming $F(0)=F(T_0)$). Clearly, the free energy difference for the boron-doped graphene is less robust towards an increase of the $\mu^*$.
Next, by using the $\frac{\Delta F}{\rho(0)}$ function, the magnetic critical field can be obtained from the formula:
\begin{align}
    \frac{H_C}{\sqrt{\rho(0)}}=\sqrt{-8\pi[\Delta F/\rho(0)]}.
    \label{eq6}
\end{align}
The upper panels of the Figs. (\ref{f1}) represent thermal behavior of the critical magnetic field for the selected values of the parameter $\mu^*$. It can be observed that the $\frac{H_C}{\sqrt{\rho(0)}}$ function is decreasing with the increase of the temperature. Moreover, the value of the critical field strongly diminishes upon increase of the Coulomb pseudopotential. This fact can be seen from the ratios obtained for the electron- and hole-doping case respectively:
\begin{align}
    [H_C (T_0)]_{\mu^*=0.3}/H_C (T_0)]_{\mu^*=0.1}\approx 0.8,
    \label{eq7}
\end{align}
and
\begin{align}
    [H_C (T_0)]_{\mu^*=0.3}/H_C (T_0)]_{\mu^*=0.1}\approx 0.66.
    \label{eq8}
\end{align}
From values of these ratios, one can conclude that the superconducting state for the $h-CN$ is more stable under change of the Coulomb pseudopotential.

\begin{table*}
\centering
\caption{Values of the thermodynamic parameters of the superconducting state for the $h-$CN and $h-$CB.}
\begin{tabular*}{\textwidth}{@{\extracolsep{\stretch{1}}}*{3}{l}@{}}
\toprule
\hline
& $h-$CN & $h-$CB \\
\hline
\midrule
$\lambda$ & 3.35 & 1.34 \\
$\omega_{\rm{max}}$ (meV) & 132.6 & 124.5 \\ 
$\omega_{c}$ (meV) & $10 \omega_{max}$ & $10 \omega_{max}$ \\
$\mu^{\star}$ & $\left< 0.1, 0.2, 0.3 \right>$ & $\left< 0.1, 0.2, 0.3 \right>$ \\
$T_{C}$ (K) & $\left< 105.6, 92.2, 83.8 \right>$ & $\left< 55.9, 44.9, 38.1 \right>$ \\
$2\Delta(0)$ (meV) & $\left< 47.95, 41.36, 37.22 \right>$ & $\left< 20.59, 16.09, 13.43 \right>$ \\
$R_{\Delta}$ & $\left< 5.27, 5.21, 5.16 \right>$ & $\left< 4.27, 4.16, 4.09 \right>$ \\
$R_{C}$ & $\left< 2.87, 2.72, 2.28 \right>$ & $\left< 2.41, 2.39, 1.74 \right>$ \\
$R_{H}$ & $\left< 0.128, 0.125, 0.125 \right>$ & $\left< 0.138, 0.140, 0.144 \right>$ \\
\hline
\bottomrule                             
\end{tabular*}
    \label{tab1}
\end{table*}

Accordingly, the thermal characteristics of the superconducting phase $C^S$ from the difference in the specific heat between superconducting and normal state ($\Delta C=C^S-C^N$) takes form:
\begin{align}
    \frac{\Delta C}{k_B \rho(0)}=-\frac{1}{\beta}\frac{d^2|\Delta F/ \rho(0)|}{d(k_B T)^2},
    \label{eq9}
\end{align}
where the specific heat of the normal state has been obtained from the formula:
\begin{align}
    \frac{C^N}{k_B \rho(0)}=\frac{\gamma}{\beta},
    \label{eq10}
\end{align}
with the Sommerfeld constant given as $\gamma \equiv\frac{2}{3}\pi^2(1+\lambda)$. The detailed derivation of equations (\ref{eq9}) and (\ref{eq10}) can be seen in \cite{blezius1987,carbotte1990}. In the Fig. (\ref{f2}) we have presented thermal behavior of the specific heat $C^N$ and $C^S$ for the $h-$CN and $h-$CB materials. Both functions increase with the increase of the temperature: for the superconducting state $C^S$ changes exponentially at low temperatures, while for the normal state, increase of the values is linear. Moreover, the specific heat of the superconducting phase is also affected by the increase of the Coulomb pseudopotential as can be seen from the analysis for $\mu^*\in\{0.1,0.2,0.3\}$. Is it worth to note, that characteristic jump in the $C^S$ occurs for the $T=T_C$ and the value of the specific heat jump is lowered by the depairing electron correlations in the $h-$CN and $h-$CB structures. In fact:
\begin{align}
    [\Delta C (T_C)]_{\mu^*=0.3}/[\Delta C (T_C)]_{\mu^*=0.1}\approx 0.67,
    \label{eq11}
\end{align}
and
\begin{align}
    [\Delta C (T_C)]_{\mu^*=0.3}/[\Delta C (T_C)]_{\mu^*=0.1}\approx 0.55,
    \label{eq12}
\end{align}
for the electron- and hole-doped graphene, respectively. It is important to notice, that the described behavior matches expected characteristics for the phonon-mediated superconductivity, as stated in \cite{carbotte1990}. Hence, it can be stated that the results presented in Figs. (\ref{f1}) and (\ref{f2}) confirm the electron-phonon character of the pairing mechanism for the discussed graphene structures.

To this end, we note that our previous analysis allow us to calculate the dimensionless thermodynamic parameters:
\begin{align}
R_H=\frac{T_C C^N(T_C)}{H_C^2},\;\;\; R_C=\frac{\Delta C(T_C)}{C^N (T_C)},
\end{align}
and,
\begin{align}
    R_{\Delta}=\frac{2 \Delta(0)}{k_B T_C},
\end{align}
which values are presented in the Table (\ref{tab1}). The estimated values are different from their counterparts obtained within the BCS theory: $[R_H]_{BCS}=0.168$, $[R_C]_{BCS}=1.43$ and $[R_\Delta]_{BCS}=3.53$ \cite{bardeen1957,bardeen1957b}. Therefore, analysis presented here suggest the pivotal role of the retardation and strong coupling effects in the considered graphene structures.

\section{Summary and conclusions}

In the following work, we have extended previous investigations of the superconducting state in the $h-$CN and $h-$CB allotropes. We have conducted our analysis within the Migdal-Eliashberg formalism in order to account for the strong-coupling and phonon-mediated character of superconducting phase in these graphene structures. To be specific, the presented analysis involved description of the thermodynamic critical field, the free energy difference and the specific heat for the superconducting state in both scenarios. In order to be as general as possible, our analysis have been performed for the three different values of the Coulomb pseudopotential $\mu^*\in \{0.1,0.2,0.3\}$. It was shown that doping graphene with electrons seems to be more favorable in case of enhancing its superconducting properties. Moreover, our analysis confirmed the strong-coupling behavior of the considered materials.

\bibliographystyle{apsrev}
\bibliography{bib2}

\begin{thebibliography}{20}
\expandafter\ifx\csname natexlab\endcsname\relax\def\natexlab#1{#1}\fi
\expandafter\ifx\csname bibnamefont\endcsname\relax
  \def\bibnamefont#1{#1}\fi
\expandafter\ifx\csname bibfnamefont\endcsname\relax
  \def\bibfnamefont#1{#1}\fi
\expandafter\ifx\csname citenamefont\endcsname\relax
  \def\citenamefont#1{#1}\fi
\expandafter\ifx\csname url\endcsname\relax
  \def\url#1{\texttt{#1}}\fi
\expandafter\ifx\csname urlprefix\endcsname\relax\def\urlprefix{URL }\fi
\providecommand{\bibinfo}[2]{#2}
\providecommand{\eprint}[2][]{\url{#2}}

\bibitem[{\citenamefont{Novoselov et~al.}(2004)\citenamefont{Novoselov, Geim,
  Morozov, Jiang, Zhang, Dubonos, Grigorieva, and Firsov}}]{novoselov2004}
\bibinfo{author}{\bibfnamefont{K.~S.} \bibnamefont{Novoselov}},
  \bibinfo{author}{\bibfnamefont{A.~K.} \bibnamefont{Geim}},
  \bibinfo{author}{\bibfnamefont{S.~V.} \bibnamefont{Morozov}},
  \bibinfo{author}{\bibfnamefont{D.}~\bibnamefont{Jiang}},
  \bibinfo{author}{\bibfnamefont{Y.}~\bibnamefont{Zhang}},
  \bibinfo{author}{\bibfnamefont{S.~V.} \bibnamefont{Dubonos}},
  \bibinfo{author}{\bibfnamefont{I.~V.} \bibnamefont{Grigorieva}},
  \bibnamefont{and} \bibinfo{author}{\bibfnamefont{A.~A.}
  \bibnamefont{Firsov}}, \bibinfo{journal}{Science}
  \textbf{\bibinfo{volume}{306}}, \bibinfo{pages}{666} (\bibinfo{year}{2004}).

\bibitem[{\citenamefont{Das et~al.}(2008)\citenamefont{Das, Pisana,
  Chakraborty, Piscanec, Saha, Waghmare, Novoselov, Krishnamurthy, Geim,
  Ferrari et~al.}}]{das2008}
\bibinfo{author}{\bibfnamefont{A.}~\bibnamefont{Das}},
  \bibinfo{author}{\bibfnamefont{S.}~\bibnamefont{Pisana}},
  \bibinfo{author}{\bibfnamefont{B.}~\bibnamefont{Chakraborty}},
  \bibinfo{author}{\bibfnamefont{S.}~\bibnamefont{Piscanec}},
  \bibinfo{author}{\bibfnamefont{S.~K.} \bibnamefont{Saha}},
  \bibinfo{author}{\bibfnamefont{U.~V.} \bibnamefont{Waghmare}},
  \bibinfo{author}{\bibfnamefont{K.~S.} \bibnamefont{Novoselov}},
  \bibinfo{author}{\bibfnamefont{H.~R.} \bibnamefont{Krishnamurthy}},
  \bibinfo{author}{\bibfnamefont{A.~K.} \bibnamefont{Geim}},
  \bibinfo{author}{\bibfnamefont{A.~C.} \bibnamefont{Ferrari}},
  \bibnamefont{et~al.}, \bibinfo{journal}{Nat. Nanotech.}
  \textbf{\bibinfo{volume}{3}}, \bibinfo{pages}{210} (\bibinfo{year}{2008}).

\bibitem[{\citenamefont{Ye et~al.}(2010)\citenamefont{Ye, Inoue, Kobayashi,
  Kasahara, Yuan, Shimotani, and Iwasa}}]{Ye2010}
\bibinfo{author}{\bibfnamefont{J.}~\bibnamefont{Ye}},
  \bibinfo{author}{\bibfnamefont{S.}~\bibnamefont{Inoue}},
  \bibinfo{author}{\bibfnamefont{K.}~\bibnamefont{Kobayashi}},
  \bibinfo{author}{\bibfnamefont{Y.}~\bibnamefont{Kasahara}},
  \bibinfo{author}{\bibfnamefont{H.~T.} \bibnamefont{Yuan}},
  \bibinfo{author}{\bibfnamefont{H.}~\bibnamefont{Shimotani}},
  \bibnamefont{and} \bibinfo{author}{\bibfnamefont{Y.}~\bibnamefont{Iwasa}},
  \bibinfo{journal}{Nat. Mater.} \textbf{\bibinfo{volume}{9}},
  \bibinfo{pages}{125} (\bibinfo{year}{2010}).

\bibitem[{\citenamefont{Profeta et~al.}(2012)\citenamefont{Profeta, Calandra,
  and Mauri}}]{Profeta2012}
\bibinfo{author}{\bibfnamefont{G.}~\bibnamefont{Profeta}},
  \bibinfo{author}{\bibfnamefont{M.}~\bibnamefont{Calandra}}, \bibnamefont{and}
  \bibinfo{author}{\bibfnamefont{F.}~\bibnamefont{Mauri}},
  \bibinfo{journal}{Nat. Phys.} \textbf{\bibinfo{volume}{8}},
  \bibinfo{pages}{131} (\bibinfo{year}{2012}).

\bibitem[{\citenamefont{Savini et~al.}(2010)\citenamefont{Savini, Ferrari, and
  Giustino}}]{salvini2010}
\bibinfo{author}{\bibfnamefont{G.}~\bibnamefont{Savini}},
  \bibinfo{author}{\bibfnamefont{A.~C.} \bibnamefont{Ferrari}},
  \bibnamefont{and} \bibinfo{author}{\bibfnamefont{F.}~\bibnamefont{Giustino}},
  \bibinfo{journal}{Phys. Rev. Lett.} \textbf{\bibinfo{volume}{105}},
  \bibinfo{pages}{037002} (\bibinfo{year}{2010}).

\bibitem[{\citenamefont{Zhou et~al.}(2015)\citenamefont{Zhou, Sun, Wang, and
  Jena}}]{zhou2015}
\bibinfo{author}{\bibfnamefont{J.}~\bibnamefont{Zhou}},
  \bibinfo{author}{\bibfnamefont{Q.}~\bibnamefont{Sun}},
  \bibinfo{author}{\bibfnamefont{Q.}~\bibnamefont{Wang}}, \bibnamefont{and}
  \bibinfo{author}{\bibfnamefont{P.}~\bibnamefont{Jena}},
  \bibinfo{journal}{Phys. Rev. B} \textbf{\bibinfo{volume}{92}},
  \bibinfo{pages}{064505} (\bibinfo{year}{2015}).

\bibitem[{\citenamefont{Black-Schaffer and Doniach}(2007)}]{blackschaffer2007}
\bibinfo{author}{\bibfnamefont{A.~M.} \bibnamefont{Black-Schaffer}}
  \bibnamefont{and} \bibinfo{author}{\bibfnamefont{S.}~\bibnamefont{Doniach}},
  \bibinfo{journal}{Phys. Rev. B} \textbf{\bibinfo{volume}{75}},
  \bibinfo{pages}{134512} (\bibinfo{year}{2007}).

\bibitem[{\citenamefont{Kiesel et~al.}(2012)\citenamefont{Kiesel, Platt, Hanke,
  Abanin, and Thomale}}]{kiesel2012}
\bibinfo{author}{\bibfnamefont{M.~L.} \bibnamefont{Kiesel}},
  \bibinfo{author}{\bibfnamefont{C.}~\bibnamefont{Platt}},
  \bibinfo{author}{\bibfnamefont{W.}~\bibnamefont{Hanke}},
  \bibinfo{author}{\bibfnamefont{D.}~\bibnamefont{Abanin}}, \bibnamefont{and}
  \bibinfo{author}{\bibfnamefont{R.}~\bibnamefont{Thomale}},
  \bibinfo{journal}{Phys. Rev. B} \textbf{\bibinfo{volume}{86}},
  \bibinfo{pages}{020507} (\bibinfo{year}{2012}).

\bibitem[{\citenamefont{Ma et~al.}(2014)\citenamefont{Ma, Yang, Yao, and
  Lin}}]{ma2014}
\bibinfo{author}{\bibfnamefont{T.}~\bibnamefont{Ma}},
  \bibinfo{author}{\bibfnamefont{F.}~\bibnamefont{Yang}},
  \bibinfo{author}{\bibfnamefont{H.}~\bibnamefont{Yao}}, \bibnamefont{and}
  \bibinfo{author}{\bibfnamefont{H.-Q.} \bibnamefont{Lin}},
  \bibinfo{journal}{Phys. Rev. B} \textbf{\bibinfo{volume}{90}},
  \bibinfo{pages}{245114} (\bibinfo{year}{2014}).

\bibitem[{\citenamefont{Ludbrook et~al.}(2015)\citenamefont{Ludbrook, Levy,
  Nigge, Zonno, Schneider, Dvorak, Veenstra, Zhdanovich, Wong, Dosanjh
  et~al.}}]{ludbrook2015}
\bibinfo{author}{\bibfnamefont{B.~M.} \bibnamefont{Ludbrook}},
  \bibinfo{author}{\bibfnamefont{G.}~\bibnamefont{Levy}},
  \bibinfo{author}{\bibfnamefont{P.}~\bibnamefont{Nigge}},
  \bibinfo{author}{\bibfnamefont{M.}~\bibnamefont{Zonno}},
  \bibinfo{author}{\bibfnamefont{M.}~\bibnamefont{Schneider}},
  \bibinfo{author}{\bibfnamefont{D.~J.} \bibnamefont{Dvorak}},
  \bibinfo{author}{\bibfnamefont{C.~N.} \bibnamefont{Veenstra}},
  \bibinfo{author}{\bibfnamefont{S.}~\bibnamefont{Zhdanovich}},
  \bibinfo{author}{\bibfnamefont{D.}~\bibnamefont{Wong}},
  \bibinfo{author}{\bibfnamefont{P.}~\bibnamefont{Dosanjh}},
  \bibnamefont{et~al.}, \bibinfo{journal}{PNAS} \textbf{\bibinfo{volume}{112}},
  \bibinfo{pages}{11795} (\bibinfo{year}{2015}).

\bibitem[{\citenamefont{Szcz{\c{e}}{\'{s}}niak and
  Szcz{\c{e}}{\'{s}}niak}(2019)}]{szczesniak2019}
\bibinfo{author}{\bibfnamefont{D.}~\bibnamefont{Szcz{\c{e}}{\'{s}}niak}}
  \bibnamefont{and}
  \bibinfo{author}{\bibfnamefont{R.}~\bibnamefont{Szcz{\c{e}}{\'{s}}niak}},
  \bibinfo{journal}{Phys. Rev. B} \textbf{\bibinfo{volume}{99}},
  \bibinfo{pages}{224512} (\bibinfo{year}{2019}).

\bibitem[{\citenamefont{D.Szcz{\c{e}}{\'{s}}niak and
  Drzazga-Szcz{\c{e}}{\'{s}}niak}(2021)}]{szczesniak2021}
\bibinfo{author}{\bibnamefont{D.Szcz{\c{e}}{\'{s}}niak}} \bibnamefont{and}
  \bibinfo{author}{\bibfnamefont{E.~A.}
  \bibnamefont{Drzazga-Szcz{\c{e}}{\'{s}}niak}}, \bibinfo{journal}{EPL}
  \textbf{\bibinfo{volume}{135}}, \bibinfo{pages}{67002}
  (\bibinfo{year}{2021}).

\bibitem[{\citenamefont{Migdal}(1958)}]{migdal1958}
\bibinfo{author}{\bibfnamefont{A.~B.} \bibnamefont{Migdal}},
  \bibinfo{journal}{Sov. Phys. JETP} \textbf{\bibinfo{volume}{34 (7)}},
  \bibinfo{pages}{996} (\bibinfo{year}{1958}).

\bibitem[{\citenamefont{Eliashberg}(1960)}]{eliashberg1960}
\bibinfo{author}{\bibfnamefont{G.~M.} \bibnamefont{Eliashberg}},
  \bibinfo{journal}{Sov. Phys. JETP} \textbf{\bibinfo{volume}{11}},
  \bibinfo{pages}{696} (\bibinfo{year}{1960}).

\bibitem[{\citenamefont{Carbotte}(1990)}]{carbotte1990}
\bibinfo{author}{\bibfnamefont{J.~P.} \bibnamefont{Carbotte}},
  \bibinfo{journal}{Rev. Mod. Phys.} \textbf{\bibinfo{volume}{62}},
  \bibinfo{pages}{1027} (\bibinfo{year}{1990}).

\bibitem[{\citenamefont{Szcz{\c{e}}{\'{s}}niak}(2006)}]{szczesniak2006}
\bibinfo{author}{\bibfnamefont{R.}~\bibnamefont{Szcz{\c{e}}{\'{s}}niak}},
  \bibinfo{journal}{Acta Phys. Polon. A} \textbf{\bibinfo{volume}{109}},
  \bibinfo{pages}{179} (\bibinfo{year}{2006}).

\bibitem[{\citenamefont{Ashcroft}(2004)}]{ashcroft2004}
\bibinfo{author}{\bibfnamefont{N.}~\bibnamefont{Ashcroft}},
  \bibinfo{journal}{Phys. Rev. Lett.} \textbf{\bibinfo{volume}{92}},
  \bibinfo{pages}{187002} (\bibinfo{year}{2004}).

\bibitem[{\citenamefont{Blezius and Carbotte}(1987)}]{blezius1987}
\bibinfo{author}{\bibfnamefont{J.}~\bibnamefont{Blezius}} \bibnamefont{and}
  \bibinfo{author}{\bibfnamefont{J.~P.} \bibnamefont{Carbotte}},
  \bibinfo{journal}{Phys. Rev. B} \textbf{\bibinfo{volume}{36}},
  \bibinfo{pages}{3622} (\bibinfo{year}{1987}).

\bibitem[{\citenamefont{Bardeen
  et~al.}(1957{\natexlab{a}})\citenamefont{Bardeen, Cooper, and
  Schrieffer}}]{bardeen1957}
\bibinfo{author}{\bibfnamefont{J.}~\bibnamefont{Bardeen}},
  \bibinfo{author}{\bibfnamefont{L.~N.} \bibnamefont{Cooper}},
  \bibnamefont{and} \bibinfo{author}{\bibfnamefont{J.~R.}
  \bibnamefont{Schrieffer}}, \bibinfo{journal}{Phys. Rev.}
  \textbf{\bibinfo{volume}{106}}, \bibinfo{pages}{162}
  (\bibinfo{year}{1957}{\natexlab{a}}).

\bibitem[{\citenamefont{Bardeen
  et~al.}(1957{\natexlab{b}})\citenamefont{Bardeen, Cooper, and
  Schrieffer}}]{bardeen1957b}
\bibinfo{author}{\bibfnamefont{J.}~\bibnamefont{Bardeen}},
  \bibinfo{author}{\bibfnamefont{L.~N.} \bibnamefont{Cooper}},
  \bibnamefont{and} \bibinfo{author}{\bibfnamefont{J.~R.}
  \bibnamefont{Schrieffer}}, \bibinfo{journal}{Phys. Rev.}
  \textbf{\bibinfo{volume}{108}}, \bibinfo{pages}{1175}
  (\bibinfo{year}{1957}{\natexlab{b}}).

\end{thebibliography}
\end{document}